# Substrate-regulated morphology of graphene

Teng Li[*], Zhao Zhang

*Department of Mechanical Engineering, University of Maryland, College Park, MD 20742*

*Maryland NanoCenter, University of Maryland, College Park, MD 20742*

**Abstract**

We delineate a general theoretical framework to determine the substrate-regulated graphene morphology through energy minimization. We then apply such a framework to study the graphene morphology on a substrate with periodic surface grooves. Depending on the substrate surface roughness and the graphene-substrate interfacial bonding energy, the equilibrium morphology of graphene ranges from 1) closely conforming to the substrate, to 2) remaining flat on the substrate. Interestingly, in certain cases, the graphene morphology snaps between the above two limiting states. Our quantitative results envision a promising strategy to precisely control the graphene morphology over large areas. The rich features of the substrate-regulated graphene morphology (e.g., the snap-through instability) can potentially lead to new design concepts of functional graphene device components.

[*] Corresponding author. Email: LiT@umd.edu

# 1. Introduction

The discovery of graphene in 2004 sparked a surge of scientific and technological interest [1-5]. A single layer of carbon atoms densely packed in a honeycomb crystal lattice, graphene exhibits unusual properties, such as ultra-high intrinsic mobility[6, 7] and intrinsic strength[8]. These exceptional properties have led to the emergence of a new paradigm of materials science and condensed-matter physics [2, 5], and have also inspired an array of tantalizing potential applications[9-13], ranging from flexible and invisible displays to chemical and biochemical sensing arrays. Enthusiasm for graphene-based applications aside, there are substantial challenges to the realization of such applications. One significant challenge is to precisely control the morphology of graphene over large areas. Graphene is intrinsically non-flat and tends to be corrugated due to the instability of two-dimensional crystals[14, 15]. The corrugation in freestanding graphene forms spontaneously, owing to thermal fluctuation, and therefore is random[15]. The corrugating physics of freestanding graphene is closely tied to its electronic properties[16, 17]. The random morphology of graphene can lead to unpredictable electronic properties, thus resulting in unstable performance of graphene devices. Therefore, controlling the graphene morphology over large areas is crucial in enabling future graphene-based applications. This paper studies the morphology of graphene regulated by the surface features of an underlying substrate. The quantitative results from the present study demonstrate a promising approach to achieving precise control of graphene morphology over large areas.

The existence of freestanding graphene has been attributed to their random intrinsic corrugations: the out-of-plane corrugations lead to increased strain energy but stabilize the random thermal fluctuation [14, 15, 18]. Graphene supported by a substrate (e.g., $SiO_2$) also corrugates, which is often attributed to graphene's intrinsic corrugations and is speculated to be relevant to graphene device performance[16, 19, 20]. Recent experiments, however, showed that the graphene corrugation on a $SiO_2$ substrate results from the mechanical interaction between the graphene and the $SiO_2$ surface[19, 21]. Via a combined-SEM/AFM/STM technique, atomic-resolution images of the graphene on $SiO_2$ in real space clearly revealed that the graphene partially conforms to the underlying $SiO_2$ substrate, and is about 60% smoother than the $SiO_2$ surface[19]. It has been further confirmed that graphene and few-layer graphene also partially follow the surface morphology of various substrates (e.g., GaAs, InGaAs and $SiO_2$) [20-23]. The regulated extrinsic corrugations in substrate-supported graphene are essentially distinct from the random intrinsic corrugations in freestanding graphene. The extrinsic substrate regulation on the graphene morphology is shown to be strong enough to prevail over the intrinsic random corrugations in graphene.

So far, the available experimental evidence on the morphology of substrate-supported graphene is suggestive, but preliminary: the quantitative relationship between the graphene morphology and the substrate surface roughness has not been studied; the effect of graphene-substrate interaction on the graphene morphology remains elusive. To address these largely unexplored issues, in this paper, we delineate a theoretical framework to determine the substrate-regulated graphene morphology through energy minimization. We then apply such a framework to study the graphene morphology on a substrate with periodic surface grooves. Depending on substrate surface roughness and graphene-substrate interfacial adhesion, the equilibrium morphology of graphene ranges from 1) closely conforming to the substrate, to 2) remaining flat on the substrate. Interestingly, in certain cases, the graphene morphology snaps between the above two distinct states. The quantitative results from the present study demonstrate a promising approach to achieving precise control of graphene morphology over large areas. Since the morphology of graphene strongly influences its electronic characteristics, it will be possible to design desired graphene device components (e.g., with tunable electrical conductivity) by tailoring the graphene to the desired morphology.

## 2. Energetic framework

The equilibrium graphene morphology regulated by the underlying substrate is governed by the interplay among three types of free energies: (1) *graphene-substrate interaction energy,* (2) *graphene strain energy and* (3) *substrate strain energy*.

(1)     The interaction between mechanically-exfoliated graphene and its underlying substrate is usually weak and can be characterized by van der Waals forces. For graphene epitaxially grown from a substrate, the graphene-substrate interaction energy results from their chemical bonding. In practice, weak physical bonding (e.g., van der Waals forces) and strong chemical bonding may co-exist in the graphene-on-substrate structure. The contributions of the chemical bonding to the interaction energy is additive to that of the van de Waals bonding.

(2)     As the graphene partially conforms to the substrate surface morphology, the graphene strain energy increases, resulting from the out-of-plane bending as well as the in-plane stretching. Furthermore, the graphene out-of-plane deformation defines its resulting morphology.

(3)     The substrate strain energy depends on the substrate stiffness and the external mechanical loads. Graphene has been fabricated mostly on rigid substrates (e.g., $SiO_2$). Without external mechanical loads, the interaction between the ultra-thin graphene and the rather thick substrate results in negligible strain energy in the substrate. If the graphene, however, is transferred onto a flexible substrate (e.g., polymers or elastomers)[13, 24], and the resulting structure is subject to large deformation, the strain energy of the substrate can become comparable to that of the graphene and the graphene-substrate interaction energy, and thus needs to be considered to determine the equilibrium graphene morphology.

The graphene partially conforming to an underlying substrate can then be understood as follows. As the graphene corrugates to follow the substrate surface morphology, the graphene strain energy and the substrate strain energy increases; on the other hand, by partially conforming to the substrate, the graphene-substrate interaction energy decreases. The total free energy of the system (denoted by the sum of the graphene strain energy and the graphene-substrate interaction energy) minimizes, from which the equilibrium graphene morphology on the substrate can be determined (Fig. 1).

### 3. Graphene morphology regulated by substrate surface grooves: A case study

In this section, to benchmark the above energetic framework, we study the graphene morphology regulated by a substrate with periodic surface grooves. Here we focus on the mechanically-exfoliated graphene spontaneously regulated by a rigid $SiO_2$ substrate, without external mechanical loads. As discussed above, the graphene-substrate interaction can be characterized by the weak van der Waals forces. The resulting strain energy in the substrate is negligible.

*Model configuration*

The substrate surface grooves lie in *y* direction and have a sinusoidal profile in *x-z* plane (Fig. 2a). A blanket graphene monolayer mechanically exfoliated on such a substrate partially conforms to the substrate surface, thus assumes a corrugated morphology similar to the substrate surface grooves but with a smaller amplitude. The graphene morphology and the substrate surface are described by

$$w_g(x) = A_g \cos\frac{2\pi x}{\lambda}$$
$$w_s(x) = A_s \cos\frac{2\pi x}{\lambda} - h \tag{1}$$

respectively, where $\lambda$ is the groove wavelength, $h$ is the distance between the middle planes of the graphene and the substrate surface, $A_g$ and $A_s$ are the amplitudes of the graphene corrugation and the substrate surface grooves, respectively (Fig. 2b). Given the symmetry and periodicity of the structure in Fig. 2, we only need to consider a graphene segment of a half sinusoidal period (e.g., $0 < x < \lambda/2$) and the underlying substrate.

*Graphene-substrate interaction energy*

The graphene-substrate interaction energy is given by summing up all interaction energies due to van der Waals force between the carbon atoms in the graphene and the substrate atoms. Denoting the interaction energy potential between a graphene-substrate atomic pair of distance $r$ by $V(r)$, the interaction energy $E_{int}$ between a graphene of area $S$ and a substrate of volume $V_s$ can be given by

$$E_{int} = \int_S \int_{V_s} V(r) \rho_s dV_s \rho_C dS \tag{2}$$

where $\rho_C$ is the homogenized carbon atom area density of graphene that is related to the equilibrium carbon-carbon bond length $l$ by $\rho_C = 4/(3\sqrt{3}l^2)$, and $\rho_s$ is the volume density of substrate atoms (i.e., the number of substrate atoms over a volume $dV_s$ is $\rho_s dV_s$) [25, 26]. It has been shown that the homogenized description of the discrete carbon atoms in graphene can capture the feature of the graphene-substrate interface[26].

The distance between a point $(x_g, 0, w_g)$ on the graphene and a point $(x_s, y_s, z_s)$ in the substrate is $r = \sqrt{(x_g - x_s)^2 + y_s^2 + (w_g - z_s)^2}$, where $w_g = A_g \cos\left(\frac{2\pi}{\lambda} x_g\right)$ and $z_s \leq A_s \cos\left(\frac{2\pi}{\lambda} x_s\right) - h$. The graphene-substrate interaction energy per unit area over a half period of graphene is given by

$$E_{int} = \frac{1}{\lambda/2} \int_0^{\lambda/2} \rho_C dx_g \int_{V_s} V(r) \rho_s dV_s$$

$$= \frac{\rho_C \rho_s}{\lambda/2} \int_0^{\lambda/2} dx_g \int_{-\infty}^{\infty} dx_s \int_{-\infty}^{\infty} dy_s \int_{-\infty}^{A_s \cos\left(\frac{2\pi}{\lambda} x_s\right) - h} V\left(\sqrt{(x_g - x_s)^2 + y_s^2 + \left(A_g \cos\left(\frac{2\pi}{\lambda} x_g\right) - z_s\right)^2}\right) dz_s \quad (3)$$

While Eq. (3) is applicable to any pair potential $V(r)$, here we use the Lennard-Jones 6-12 potential, $V_{LJ}(r) = 4\varepsilon(\sigma^{12}/r^{12} - \sigma^6/r^6)$, to represent the graphene-substrate van der Waals force, where $\sqrt[6]{2}\sigma$ is the equilibrium distance of a graphene-substrate atomic pair and $\varepsilon$ is the bonding energy at the equilibrium distance.

The van der Waals force rapidly decays as the distance of an atom pair increases from its equilibrium value. Therefore, the interaction energy between a carbon atom in the graphene and the underlying substrate can be estimated by the van de Waals interactions of this carbon atom with its adjacent portion of the substrate (e.g., within a certain distance). The interaction energy defined in Eq. (3) is then computed using a Monte Carlo numerical strategy as described below. For a given carbon atom in the graphene, only the substrate portion within a distance of $R$ to this carbon atom are taken into account in computing the graphene-substrate interaction energy (Fig. 2b). Then $n$ locations in such a substrate portion are randomly generated. The interaction energy between the given carbon atom and the substrate is then estimated by

$$E(x_g) = \rho_s V_\Pi \frac{1}{n} \sum_{i=1}^{n} V_{LJ}(r_i) \tag{4}$$

where $x_g$ is the $x$ coordinate of the given carbon atom, $r_i$ is the distance between the given carbon atom and the $i^{th}$ random substrate location, and $V_\Pi$ is the volume of the substrate portion inside a sphere with its center at the given carbon atom and a radius of $R$. The graphene-substrate interaction energy per unit area over a half period of graphene can then be estimated by

$$E_{int} = \frac{\rho_g}{\lambda/2} \int_0^{\lambda/2} E(x_g) dx_g . \tag{5}$$

As $n$ and $R$ become larger, the values of Eq. (5) converge to the theoretical value of Eq. (5). In all simulations, we take $R = 3 nm$ and $n = 10^6$, which lead to less than one per cent variation of the estimated values of $E_{int}$.

*Graphene strain energy*

As the graphene spontaneously follows the surface morphology of the substrate (imagine a fabric conforms to a corrugated surface), the strain energy in the graphene mainly results from out-of-plane bending of the graphene, while the contribution from in-plane stretching of the graphene to the strain energy is negligible. Denoting the out-of-plane displacement of the graphene by $w_g(x, y)$, the strain energy $E_g$ of the graphene over its area $S$ can be given by

$$E_g = \int_S \frac{D}{2} \left[ \left( \frac{\partial^2 w_g}{\partial x^2} + \frac{\partial^2 w_g}{\partial y^2} \right)^2 - 2(1-\nu) \left( \frac{\partial^2 w_g}{\partial x^2} \frac{\partial^2 w_g}{\partial y^2} - \left( \frac{\partial^2 w_g}{\partial x \partial y} \right)^2 \right) \right] dS \tag{6}$$

where $D$ and $\nu$ are the bending rigidity and the Poisson's ratio of graphene, respectively[27].

By substituting $w_g(x)$ defined in Eq. (1) into Eq. (6), the graphene bending energy per unit area over such a half period is given by

$$E_g = \frac{1}{\lambda/2}\int_0^{\lambda/2} \frac{D}{2}\left(\frac{\partial^2 w_g}{\partial x^2}\right)^2 dx = \frac{4\pi^4 D A_g^2}{\lambda^4}. \tag{7}$$

The equilibrium morphology of the graphene on the substrate, described by $w(x,y)$, can then be determined by minimizing the total free energy $(E_g + E_{int})$.

For a given substrate surface morphology (i.e., $\lambda$ and $A_s$), the graphene bending energy $E_g$ is a function of the amplitude of graphene corrugation $A_g$, and monotonically increases as $A_g$ increases (e.g., Eq. (4)). On the other hand, the graphene-substrate interaction energy $E_{int}$ is a function of $A_g$ and $h$. Due to the nature of van der Waals interaction, $E_{int}$ minimizes at finite values of $A_g$ and $h$. As a result, there exists a minimum value of $(E_g + E_{int})$ where $A_g$ and $h$ define the equilibrium morphology of the graphene on the substrate (Fig. 1). In simulations, the equilibrium values of $A_g$ and $h$ are obtained numerically by minimizing the sum of $E_{int}$ (from Eq. (5)) and $E_g$ (from Eq. (7)).

## 4. Results and discussion

We will next describe the simulation results using the following dimensionless groups: $A_g/A_s$, $h/\sigma$, $\lambda/A_s$, $D/\varepsilon$, and $(E_g + E_{int})A_s/D$. In all simulations, we take $D = 1.41\,eV$, $\rho_C = 3.82\times10^{19}/m^2$, $\rho_s = 6.61\times10^{28}$, $\sigma = 0.38\,nm$, and $A_s = 0.5\,nm$. These values are representative of a graphene-on-SiO$_2$ material system[25, 28-30]. Various values of $D/\varepsilon$ (i.e., 25~2000) and $\lambda/A_s$ (i.e., 1~30) are used to study the effects of interfacial bonding energy and substrate surface roughness. The simulations were conducted by running a parallel computer code through a multi-node high performance computing cluster.

Figures 3a and 3b show the normalized equilibrium amplitude of the graphene corrugation $A_g/A_s$ and the normalized equilibrium graphene-substrate distance $h/\sigma$ as a function of $D/\varepsilon$ for various $\lambda/A_s$, respectively. For a given substrate surface roughness (i.e., $\lambda/A_s$), if the graphene-substrate interfacial bonding energy is strong (i.e., small $D/\varepsilon$), $A_g$ tends to $A_s$, while $h$ becomes comparable to $\sigma$. In other words, the graphene closely follows the substrate surface morphology, and the equilibrium distance between the graphene and the substrate is comparable to the equilibrium atomic distance defined in the Lennard-Jones potential. By contrast, if the graphene-substrate interfacial bonding is weak (i.e., large $D/\varepsilon$), $A_g$ approaches zero, while $h$ tends to $2.1\sigma$. That is, the graphene is nearly flat and does not conform to the substrate surface. For a given interfacial bonding energy (i.e., $D/\varepsilon$), $A_g$ increases and $h$ decreases, as $\lambda/A_s$ increases.

Worth noting in Fig. 3 is that: for certain range of $\lambda/A_s$ (e.g., $\lambda/A_s$ = 4 or 10), there is a sharp transition in the equilibrium amplitude of the graphene corrugation as the interfacial bonding energy varies. Particularly, for $\lambda/A_s = 10$, $A_g/A_s$ drops from 0.86 to 0.27, when $D/\varepsilon = 1420$ (Fig. 3a). In other words, the graphene morphology snaps between two distinct states: closely conforming to the substrate surface and nearly remaining flat on the substrate surface, when the interfacial bonding energy reaches a threshold value. Such a snap-through instability of the graphene morphology on the substrate is also evident in Fig. 3b.

Figure 4 provides the energetic understanding of the above snap-through instability. For $\lambda/A_s = 10$, when the interfacial bonding energy is low (e.g., $D/\varepsilon$ =1250), ($E_g + E_{int}$) minimizes at $A_g/A_s$ =0.19. As $D/\varepsilon$ increases, ($E_g + E_{int}$) vs. $A_g/A_s$ curve assumes a double-well shape.

At a threshold value of $D/\varepsilon = 1420$, $(E_g + E_{int})$ minimizes at both $A_g/A_s = 0.86$ and 0.27, corresponding to the two distinct states of the graphene morphology, respectively. For $D/\varepsilon$ higher than the threshold value, the minimum of $(E_g + E_{int})$ occurs at a larger $A_g/A_s$.

Besides the interfacial bonding energy, the substrate surface roughness also can influence the graphene morphology. Figure 5a further shows the effect of substrate surface roughness $\lambda/A_s$ on the graphene amplitude $A_g/A_s$ for various values of $D/\varepsilon$. For a given interfacial bonding energy $D/\varepsilon$, there exists a threshold $\lambda_{min}$, smaller than which $A_g/A_s = 0$ (i.e., the graphene is flat, and thus not conforming to the substrate surface); and a threshold $\lambda_{max}$, greater than which $A_g/A_s = 1$ (i.e., the graphene fully conforming to the substrate surface). As $\lambda$ increases from $\lambda_{min}$ to $\lambda_{max}$, $A_g/A_s$ ramps up from zero to one. This can be understood as follows. For a given amplitude of substrate surface groove $A_s$, if the groove wavelength $\lambda$ is small, conforming to substrate surface results in a significant increase in the graphene bending energy (e.g., $E_g \propto 1/\lambda^4$ in Eq. (7)). Consequently, $A_g$ tends to zero. On the other hand, if $\lambda$ is large, the graphene bending energy becomes negligible; the graphene closely follows the substrate morphology (i.e., $A_g/A_s$ tends to one). Figure 5b plots the equilibrium graphene-substrate distance $h/\sigma$ as a function of $\lambda/A_s$ for the case of $D/\varepsilon = 1000$. The equilibrium distance $h$ tends to be constant when $\lambda > \lambda_{max}$ or $\lambda < \lambda_{min}$, corresponding to the two distinct states described above (as illustrated in the insets of Fig. 5a). The difference in the equilibrium distance of the two distinct states is $\Delta h = 1.2\sigma = 0.459 nm$, which is well close to the amplitude of the substrate surface groove $A_s = 0.5 nm$, further demonstrating the distinction between the graphene morphologies in the two limiting states.

Also worth of noting in Fig. 5 is that, for certain range of graphene-substrate interfacial bonding energy (e.g., $D/\varepsilon > 1000$), the snap-through instability of the graphene morphology, similar to that shown in Fig. 3, exists. That is, the graphene morphology sharply switches between two distinct states: closely conforming to the substrate surface and nearly remaining flat on the substrate surface, when the substrate surface roughness $\lambda/A_s$ reaches a threshold value. Such a threshold value increases as $D/\varepsilon$ increases. The snap-through instability shown in Fig. 5 also results from the double-well feature of the total energy profile at the threshold value of $\lambda/A_s$, similar to that shown in Fig. 4.

The results reported here, combined with recent experimental observations[19, 21], reveal a promising strategy to achieve quantitative control of the graphene morphology by tailoring the surface profile of the underlying substrate. While it is difficult to directly manipulate freestanding graphene at the atomistic scale[31], it is feasible to pattern the substrate surface with nano-scale features via micro/nano- fabrication techniques[32-36]. The graphene on such a patterned substrate surface will assume a regular morphology, rather than random thermal fluctuation as in its freestanding counterpart. Such a strategy is justified by recent direct experimental observation of the suppression of any intrinsic ripples in graphene regulated by the atomically flat terraces of cleaved mica surfaces [37]. This strong experimental evidence also suggests the robustness of the snap-through instability of graphene on substrates over the thermal fluctuations. If graphene can be tailored into desired morphologies, its unusual properties (e.g., tunable electrical conductivity and mobility), which are impossible in freestanding graphene, may be achieved. These unusual mechanical and electrical properties of the graphene can be potentially used to develop graphene-based devices. For example, the snap-through instability of the graphene demonstrated in this letter can possibly enable the design of graphene switches for nano-electronics.

In our model, the graphene is assumed to adhere to the substrate surface spontaneously during fabrication and result in negligible deformation of the substrate. When a graphene-substrate laminate is subject to external loading, the graphene strain energy due to stretching and the substrate strain energy may also need to be considered to determine the graphene morphology. In this sense, the present model overestimates the equilibrium amplitude of the graphene morphology. We also assume the weak graphene/substrate interaction. In practice, it is possible to have chemical bondings or pinnings between the graphene and the substrate, leading to enhanced interfacial bonding [38-40]. In this sense, the present model underestimates the equilibrium amplitude of the graphene morphology. The contribution of the chemical bonding to the interaction energy is additive to that of the van de Waals bonding, thus can be readily incorporated in the energetic framework. We also assume that the regulated graphene morphology has the same wavelength of the substrate surface grooves. This assumption is justified if the substrate surface is modestly rough. On a severely rough substrate surface, the graphene may assume morphology of a longer wavelength to reduce the strain energy[41]. In this paper, we study the graphene morphology modulated by a substrate surface with two-dimensional features. The surface features of natural substrates are often random in three dimensions. It is also feasible to engineer the substrate surfaces with regular nano-scale three-dimensional features (e.g., patterned islands, pillars, and/or wells). The equilibrium graphene morphology on such substrate surfaces is yet to be explored. For example, the graphene on a substrate surface with bi-sinusoidal fluctuations (i.e., in both $x$ and $y$ directions) also exhibits the snap-through instability reported above, but with a smaller equilibrium amplitude[42]. Further studies are needed to address the above issues, and will be reported elsewhere.

**5. Concluding remarks**

In summary, this paper envisions a promising strategy to precisely control the graphene morphology over large areas via substrate regulation. We delineate a theoretical framework to determine the substrate-regulated morphology of the graphene through energy minimization. A case study reveals the graphene snap-through instability on a substrate with sinusoidal surface grooves. The graphene with controlled morphology could enable systematic exploration into the effect of corrugation-induced strain on the transport properties of graphene, an important but largely unexplored topic. Furthermore, the rich features of the substrate-regulated graphene morphology (e.g., the snap-through instability) could find their potential applications in designing new functional graphene device components. We therefore call for new experiments to demonstrate the above-envisioned strategy.

**Acknowledgments:**

This work has been supported by the Minta-Martin Foundation and by the Ralph E. Powe Jr. Faculty Award from Oak Ridge Associated Universities to T.L.. Z.Z. also acknowledges the support of the A. J. Clark Fellowship. The authors would like to thank E. D. Williams for helpful discussions and the anonymous referees for valuable comments.

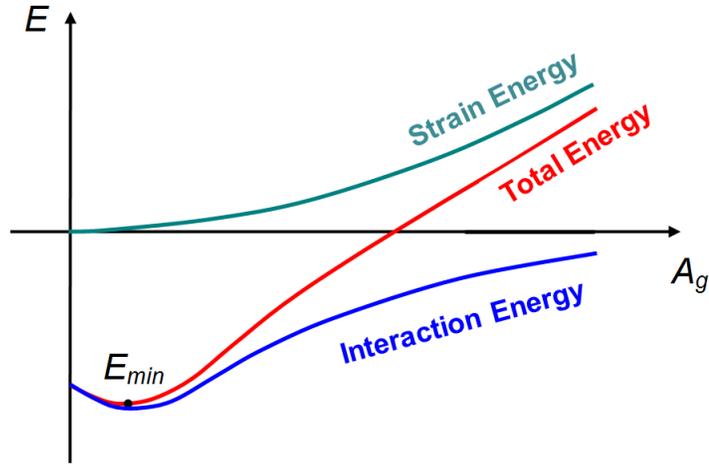

**Fig. 1**. (color online) Schematics of the energetics of the substrate regulation on graphene morphology. The strain energy and the graphene-substrate interaction energy are plotted as functions of the graphene corrugation amplitude $A_g$. The total free energy minimizes at an equilibrium value of $A_g$. A similar energy profile holds for the total energy as a function of $h$ (not shown).

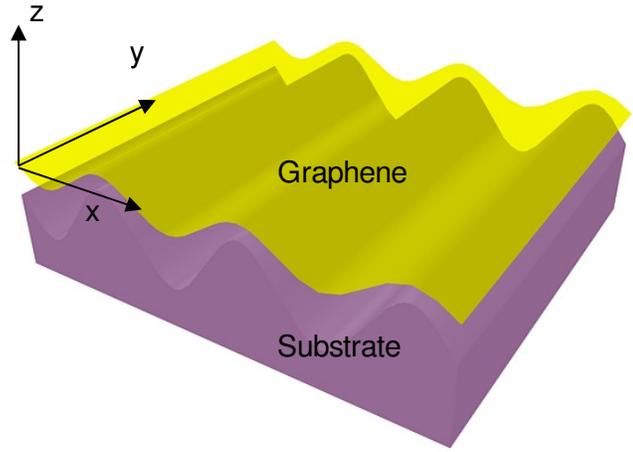

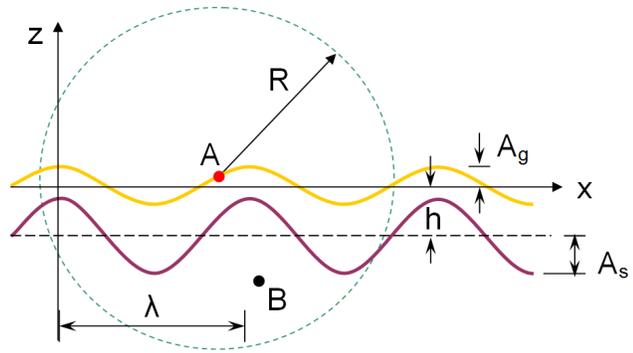

**Fig. 2.** (color online) (a) Schematics of a blanket graphene partially conforming to a substrate with sinusoidal surface grooves. (b) The view of the graphene and the substrate surface in *x-z* plane. Point A denotes a carbon atom in the graphene, and point B denotes a substrate location within a distance of *R* from point A.

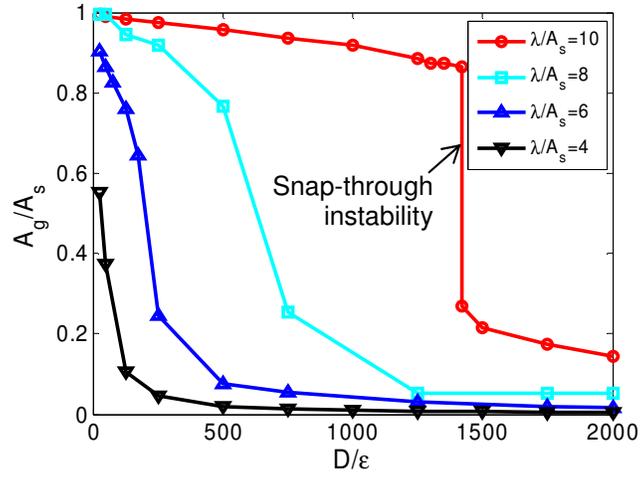

(a)

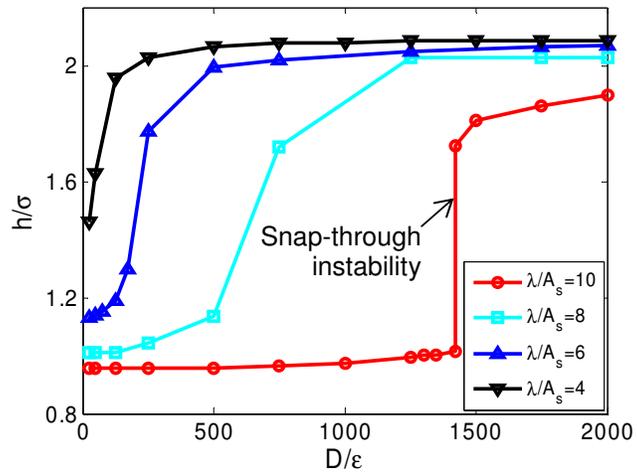

(b)

**Fig. 3.** (color online) (a) $A_g/A_s$ and (b) $h/\sigma$ as functions of $D/\varepsilon$ for various $\lambda/A_s$, respectively. For $\lambda/A_s$ =10 and $D/\varepsilon$ =1420, the equilibrium graphene morphology snaps between two distinct states: 1) closely conforming to the substrate surface and 2) nearly remaining flat on the substrate.

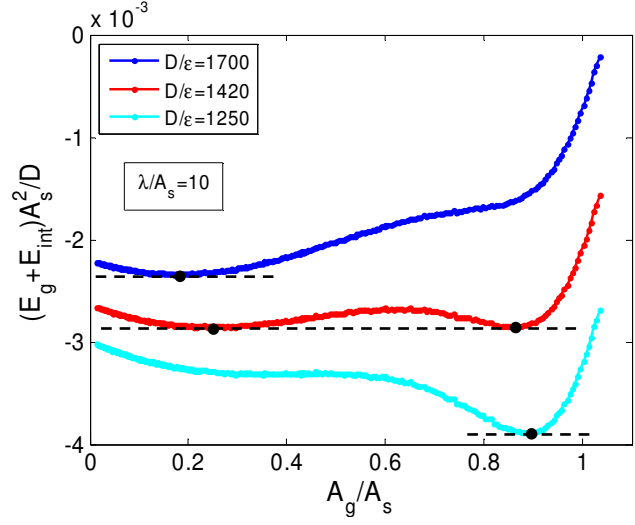

**Fig. 4.** (color online) The normalized total energy as a function of $A_g/A_s$ for various $D/\varepsilon$. At a threshold value of $D/\varepsilon = 1420$, $(E_g + E_{int})$ minimizes at both $A_g/A_s = 0.27$ and 0.86, corresponding to the two distinct states of the graphene morphology, respectively.

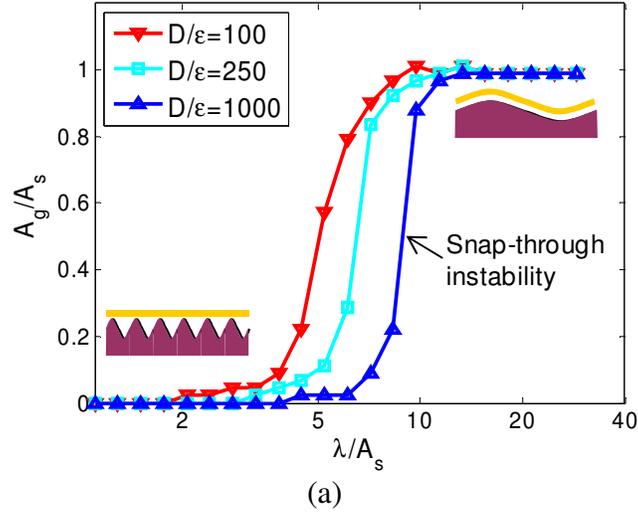

(a)

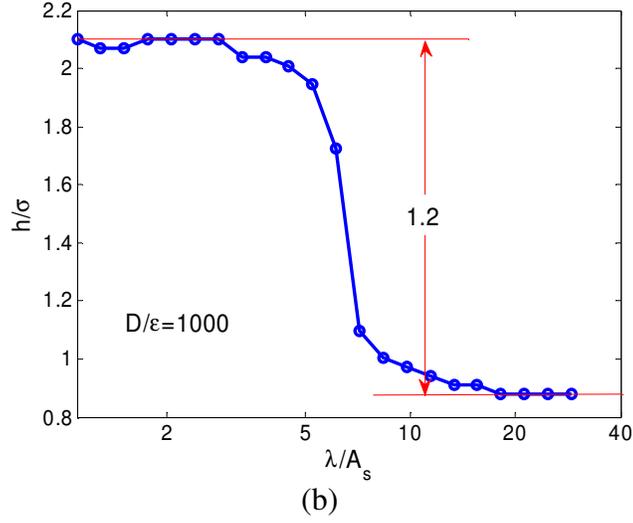

(b)

**Fig. 5.** (color online) (a) $A_g/A_s$ as a function of $\lambda/A_s$ for various $D/\varepsilon$. The insets illustrate the two distinct states of the equilibrium graphene morphology. (b) $h/\sigma$ as a function of $\lambda/A_s$ for $D/\varepsilon=1000$. The difference in the equilibrium distance of the two distinct states ($\Delta h/\sigma=1.2$) agrees well to the amplitude of the substrate surface groove ($A_s/\sigma=1.3$).